\documentclass[graybox]{svmult}

\usepackage{amssymb}
\usepackage{mathptmx}       % selects Times Roman as basic font
\usepackage{helvet}         % selects Helvetica as sans-serif font
\usepackage{courier}        % selects Courier as typewriter font
\usepackage{type1cm}        % activate if the above 3 fonts are
\usepackage{makeidx}         % allows index generation
\usepackage{graphicx}        % standard LaTeX graphics tool
\usepackage{multicol}        % used for the two-column index
\usepackage[bottom]{footmisc}% places footnotes at page bottom

\newcommand{\be}{\begin{equation}}
\newcommand{\ee}{\end{equation}}
\newcommand{\beq}{\begin{eqnarray}}
\newcommand{\eeq}{\end{eqnarray}}

\makeindex             % used for the subject index
\begin{document}

\title*{Gravitational waves from rapidly rotating neutron stars}
\author{Brynmor Haskell, Nils Andersson, Caroline D`Angelo, Nathalie Degenaar, Kostas Glampedakis, Wynn C.G. Ho, Paul D. Lasky, Andrew Melatos, Manuel Oppenoorth,  Alessandro Patruno, Maxim Priymak}
\authorrunning{Brynmor Haskell et al.}
\institute{Brynmor Haskell \at School of Physics, The University of Melbourne, Parkville, VIC 3010, Australia  \email{bhaskell@unimelb.edu.au}\\
Andrew Melatos, Paul D. Lasky, Maxim Priymak \at School of Physics, The University of Melbourne, Parkville, VIC 3010, Australia \\
Nils Andersson, Wynn C.G. Ho \at Mathematical Sciences and STAG Research Centre, University of Southampton, Southampton, SO17 1BJ, UK\\
Nathalie Degenaar \at University of Michigan, Department of Astronomy, Ann Arbor, MI 48109, USA\\
Kostas Glampedakis \at Departamento de Fisica, Universidad de Murcia, Murcia, E-30100, Spain\\
Manuel Oppenoorth \at Department of Earth Sciences, Utrecht University, Budapestlaan 4, 3584 CD, Utrecht, The Netherlands\\
Alessandro Patruno, Caroline D`Angelo \at Leiden Observatory, Leiden University, Postbus 9513, 2300 RA, Leiden, The Netherlands\\ ASTRON, the Netherlands Institute for Radio Astronomy, Postbus 2, 7990 AA, Dwingeloo, The Netherlands}
\maketitle

%\abstract*{Each chapter should be preceded by an abstract (10--15 lines long) that summarizes the content. The abstract will appear \textit{online} at \url{www.SpringerLink.com} and be available with unrestricted access. This allows unregistered users to read the abstract as a teaser for the complete chapter. As a general rule the abstracts will not appear in the printed version of your book unless it is the style of your particular book or that of the series to which your book belongs.
%Please use the 'starred' version of the new Springer \texttt{abstract} command for typesetting the text of the online abstracts (cf. source file of this chapter template \texttt{abstract}) and include them with the source files of your manuscript. Use the plain \texttt{abstract} command if the abstract is also to appear in the printed version of the book.}

\abstract{Rapidly rotating neutron stars in Low Mass X-ray Binaries have been proposed as an interesting source of gravitational waves. In this chapter we present estimates of the gravitational wave emission for various scenarios, given the (electromagnetically) observed characteristics of these systems. First of all we focus on the r-mode instability and show that a 'minimal' neutron star model (which does not incorporate exotica in the core, dynamically important magnetic fields or superfluid degrees of freedom), is not consistent with observations. We then present estimates of both thermally induced and magnetically sustained mountains in the crust. In general magnetic mountains are likely to be detectable only if the buried magnetic field of the star is of the order of $B\approx 10^{12}$ G. In the thermal mountain case we find that gravitational wave emission from persistent systems may be detected by ground based interferometers. Finally we re-asses the idea that gravitational wave emission may be balancing the accretion torque in these systems, and show that in most cases the disc/magnetosphere interaction can account for the observed spin periods.}

\section{Introduction}
\label{sec:1}

Neutron Stars (NSs) are one of the most fascinating fundamental physics laboratories in the Universe. With masses comparable to that of the sun compressed in a 10 km radius, these objects have internal densities that can easily exceed the nuclear saturation density, $\rho_0\approx 2.4 \times 10^{14}$ g/cm$^3$, allowing us to probe a regime of the strong interaction that is not accessible with terrestrial experiments. In fact, although the internal temperatures of NS can be around $T\approx 10^8 K$,  at such high densities the thermal energy of the constituents is negligible compared to their Fermi energy. Neutron Stars are thus essentially cold objects. While colliders, such as GSI at Darmstadt to the LHC at CERN, allow us to probe the high temperature regime of the QCD phase diagram (generally at low densities) and study phases such as quark gluon plasmas \cite{Brambilla}, NSs give us the opportunity to probe the high density, low temperature regime of QCD. At asymptotically high densities one expects quarks to pair in the so-called Colour-Flavour-Locked (CFL) phase \cite{alfordCFL}. At realistic NS densities, however, there is still significant uncertainty on what the ground state of matter will be, and only astrophysical observations can shed light on this fundamental problem.

In order to interpret astrophysical data it is necessary to model the interior dynamics of NSs in detail. This is a formidable task, as several complex physical processes are at work in these systems. The outer layers of the star form a crystalline crust, that can support shearing and effectively insulate the hot interior and lead to the observable electromagnetic emission \cite{Gud}. At higher densities neutrons begin to drip out of the nuclei and are expected to be superfluid. At even higher densities one has a transition to a core fluid of superfluid neutrons, superconducting protons (most likely in a type II superconducting state) and electrons. Finally, at densities higher than saturation density, the composition of the star is unknown and may include exotic particles such as hyperons or deconfined quarks. Further complications arise from the fact that in many cases one has to deal with rapid rotation (up to considerable fractions of the Keplerian breakup frequency) and with dynamically important magnetic fields (of up to $B\approx 10^{15}$ G in magnetars). The extreme compactness of NSs further complicates the problem, as the effects of General Relativity become significant, and must be taken into account.

In order to understand the different physical mechanisms at work it is thus necessary to take a multi messenger approach and combine all observational signatures of NSs. NSs are observed in several electromagnetic bands, from the radio to gamma rays, but are also likely to be interesting sources of GWs. This opens a new and exciting window, as electromagnetic radiation originates mainly from the outer layers of the star while GWs interact weakly with matter and carry a strong imprint of the physics at work in the high density interior. There are several GW emission mechanisms that involve NSs and could lead to emission at the level detectable with current and next generation ground based GW detectors, such as Advanced LIGO, Advanced Virgo Kagra or the Einstein Telescope (ET). The most promising sources are clearly NS-NS binaries, which are the prime target for Advanced LIGO and could carry the imprint of the equation of state of dense matter \cite{DelP}. In the following, however, we shall focus on a different class of sources, continuous sources. In particular we shall discuss several mechanisms that may lead to continuous GW emission from a rapidly rotating NS in a Low Mass X-ray Binary (LMXB) and assess the detectability of such a signal.

\section{Gravitational Wave emission mechanisms}
\label{sec:2}
% Always give a unique label
% and use \ref{<label>} for cross-references
% and \cite{<label>} for bibliographic references
% use \sectionmark{}
% to alter or adjust the section heading in the running head

There are several mechanisms that can lead to GW emission from a rapidly rotating NS. All of them are based on the idea that a non-axisymmetric perturbation will be dragged around by rotation and lead to assess GW emission. The types of perturbation can roughly be divided into two categories: either "mountains", i.e. deformations that are static (at least on dynamical timescales) in the frame of the star, or  hydrodynamical modes of oscillation being excited in the star.
The most natural location for a NS mountain is the crust, as the finite shear modulus of the crystalline crust offers the possibility of supporting a deformation \cite{Bild98}. Recent estimates of the breaking strain of the crust have shown that high pressure and gravity lead to a remarkably strong material \cite{chuck}, that could sustain mountains large enough to be detected ground based interferometers\cite{HasMount, Ben}.

Strong magnetic fields can also confine material and lead to deformations that could be potentially quite large in magnetars \cite{RicNew, Has08}. The situation is even more interesting in accreting systems, in which, although the magnetic field is globally much weaker than in magnetars, the accretion process can lead to material spreading equatorially and compressing the field, making it locally strong enough to sustain a sizeable mountain \cite{Andrew1,Andrew2, Max1}.

Modes of oscillation of the star can also grow to large amplitude and lead to gravitational radiation. The prime candidate for this kind of mechanism in LMXBs is the r-mode. This is a toroidal mode of oscillation for which the restoring force is the Coriolis force. To leading order in the slow-rotation approximation it is purely toroidal and the Eulerian velocity perturbation $\delta\mathbf{v}$ takes the form:
\be
\delta\mathbf{v}=\alpha\left(\frac{r}{R}\right)^l R\Omega \mathbf{Y}_{lm}^B e^{i\omega t}
\ee
where $ \mathbf{Y}_{lm}^B=[l(l+1)]^{-1/2}r \nabla\times (r\nabla Y_{lm})$ is the magnetic-type vector spherical harmonic (with $Y_{lm}$ the standard spherical harmonics), $R$ is the stellar radius and $\alpha$ the dimensionless amplitude of the mode \cite{Owen98}, $\Omega$ the rotation frequency of the star and $\omega$ the frequency of the mode.

It is of particular interest because not only is its frequency in the right range to be detected by ground based GW detectors (if the star is rotating at millisecond periods), but it is also generically unstable to GW emission \cite{Nils98, Morsink98}.  As we shall see in the following, this means that the mode can grow to large amplitudes, provided viscosity does not damp it on a shorter timescale than GW emission can drive it unstable.

\section{Low Mass X-ray Binaries}
\label{sec:3}

Before discussing the above mechanisms in detail, let us examine why NSs in LMXBs are interesting from a GW perspective. In an LMXB a compact object (a NS in the case that interests us) is accreting material from a low mass star ($M \lesssim 1 M_\odot$) which fills its Roche lobe. Matter leaves the secondary star and forms an accretion disc around the NS, eventually interacting with the magnetic field of the star and being accreted. Angular momentum is transfer to the NS, spinning it up and allowing for old, slow pulsars to be recycled to millisecond spin periods. This is believed to be the primary formation channel for Millisecond Radio Pulsars (MSRPs) \cite{Rec1}.

In this scenario, and provided that the magnetic fields of these systems are weak, one would expect the NS to spin up to its Keplerian break up frequency. The exact value of such frequency is equation of state dependent but, quite generally, is expected to be above $\approx 1.5$ kHz. This expectation is, however, not borne out by observations of both LMXBs and MSRPs. In both cases the frequency distribution appears to have a cutoff around 700 Hz, well below the breakup frequency \cite{Chak, Patruno10}. An additional mechanism thus needs to be at work to remove angular momentum from the system and halt the accretion induced spin up of the NS. A natural candidate would be the interaction of the stellar magnetosphere with the accretion disc. If the magnetic field is strong enough (above $\approx 10^8$ G) it can disrupt the accretion disc above the stellar surface. Matter is accreted at the truncation radius and transfers its angular momentum to the star. This can, however, only happen as long as matter at the truncation radius is rotating faster than the star. Once the NS spin exceeds this limit, accretion is centrifugally inhibited, and matter can either be expelled from the system \cite{Car1} or accrete unstably \cite{Car2}.

The first possibility was examined by White and Zhang \cite{WZ} , who considered the implications of assuming that the observed spin period of the LMXB is the spin equilibrium period, as set by the torque balance mechanism described above. Their conclusion was, based on the models and data available, that this is unlikely, as it would require both stronger magnetic fields that observed in MSRPs (i.e. it would require fields in the range $B\approx 10^9-10^{10}$ G or above) and an unexpected correlation between the magnetic field strength and the mass accretion rate. We shall discuss recent reassessments of this analysis later on and discuss how it may actually be an explanation for the observed spin distribution. Nevertheless the original analysis by White and Zhang \cite{WZ} led to renewed interest in GW emission as a mechanism to remove angular momentum from rapidly rotating neutron stars \cite{PP,Bild98} and to detailed analysis of the physical mechanisms (described in section \ref{sec:2}) that could lead to it.

\section{The r-mode instability}
\label{sec:4}

\begin{figure}[b]
\sidecaption
% For example, with the graphicx style use
\includegraphics[scale=.39]{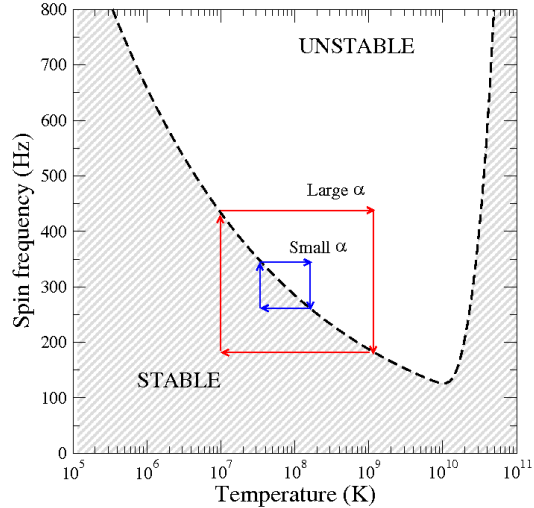}
%
% If no graphics program available, insert a blank space i.e. use
%\picplace{5cm}{2cm} % Give the correct figure height and width in cm
%
\caption{The r-mode instability window for a 1.4 $M_\odot$, $R=10$ km NS, described by an $n=1$ polytrope. We assume the 'minimal' model described in the text, with no exotica in the core. At low temperatures the main sources of damping are shear viscosity and dissipation at the crust-core boundary. At high temperatures bulk viscosity gives the main contribution. We also schematically illustrate the cycle that a system would follow in the temperature-frequency plane, both for small and large amplitude $\alpha$ of the mode.}
\label{window}
\end{figure}

Let us begin our analysis from the r-mode instability. As already mentioned the mode can only grow to large amplitudes if GW emission can drive it on a shorter timescale than viscosity can damp it. The competition between different mechanisms depends on several parameters, mainly the mass and equation of state of the star, its temperature and spin frequency. Given an equation of state (and thus a composition) for the star we can fix the mass and define an instability 'window' in the temperature vs frequency plane. In figure \ref{window}  we show the instability window for a 'minimal' neutron star model, i.e. a model in which we assume no exotica in the core, no dynamically important magnetic field or superfluid degrees of freedom, and take a simple $n=1$ polytrope as an equation of state. We consider a typical 1.4 $M_\odot$ NS with a 10 km radius and show the curve on which the driving and damping timescale are equal, i.e. the solutions of :
\be
\frac{1}{\tau_{gw}}=\sum_i \frac{1}{\tau_{V i}}
\ee
where $\tau_{gw}$ is the timescale on which GWs drive the mode unstable, which for an $n=1$ polytrope is given by \cite{NilsRev}
\be
\tau_{gw}=-47 \left(\frac{M}{1.4 M_\odot}\right)^{-1}\left(\frac{R}{10 \mbox{km}}\right)^{-4}\left(\frac{P}{1 \mbox{ms}}\right)^6 \mbox{ s},
\ee
while $\tau_{V i}$ is the viscous damping timescale for process $i$ acting in the star.
At high temperature bulk viscosity provides the main damping mechanism, while at low temperatures the main contribution is from shear viscosity, due to standard scattering processes (neutron-neutron in non-superfluid matter or electron-electron in superfluid matter \cite{PhysA}), or from viscosity at the crust-core interface.

In such a scenario an accreting neutron star, with a typical {\it core} temperature of around $10^8$ K would spin up into the unstable region due to accretion, leading to the onset of the instability. The r-mode would then rapidly grow to large amplitude, leading to fast heating. Eventually the thermal runaway is halted by neutrino emission and the star spins down due to GW emission, re-enters the stable region and cools, starting the cycle again, as depicted schematically in figure \ref{window}. The amount of heating, i.e. how far into the instability window a system can go, depends critically on the saturation amplitude for the mode, $\alpha$, as the energy dissipated by viscosity takes the form \cite{NilsRev}
\be
\frac{dE}{dt}=1.31\frac{\alpha^2\nu^2MR^2}{\tau_{sv}}
\ee
with $\tau_{sv}$ the shear viscosity damping timescale, which, for an $n=1$ polytrope and electron-electron scattering, takes the form \cite{NilsRev}:
\be
\tau_{sv}=2.2 \times 10^5 \left(\frac{M}{1.4 M_\odot}\right)^{-1}\left(\frac{R}{10\mbox{ km}}\right)^5\left(\frac{T}{10^8 \mbox{K}}\right)^2 s
\ee

If the mode can grow to large amplitudes ($\alpha\approx 1$) the system will enter well into the unstable region, but the duty cycle will be very short, less than $\approx 1\%$ \cite{Levin}. If, on the other hand, the mode saturates at a relatively low amplitude ($\alpha\approx 10^{-5}$), as calculations of non-linear couplings to other modes suggest \cite{IRA1}, the duty cycle is much longer but the system will never depart significantly from the instability curve \cite{Hyle}. In either scenario it is highly unlikely to observe a system in the unstable region.

\begin{figure}[b]
\sidecaption
% For example, with the graphicx style use
\includegraphics[scale=.42]{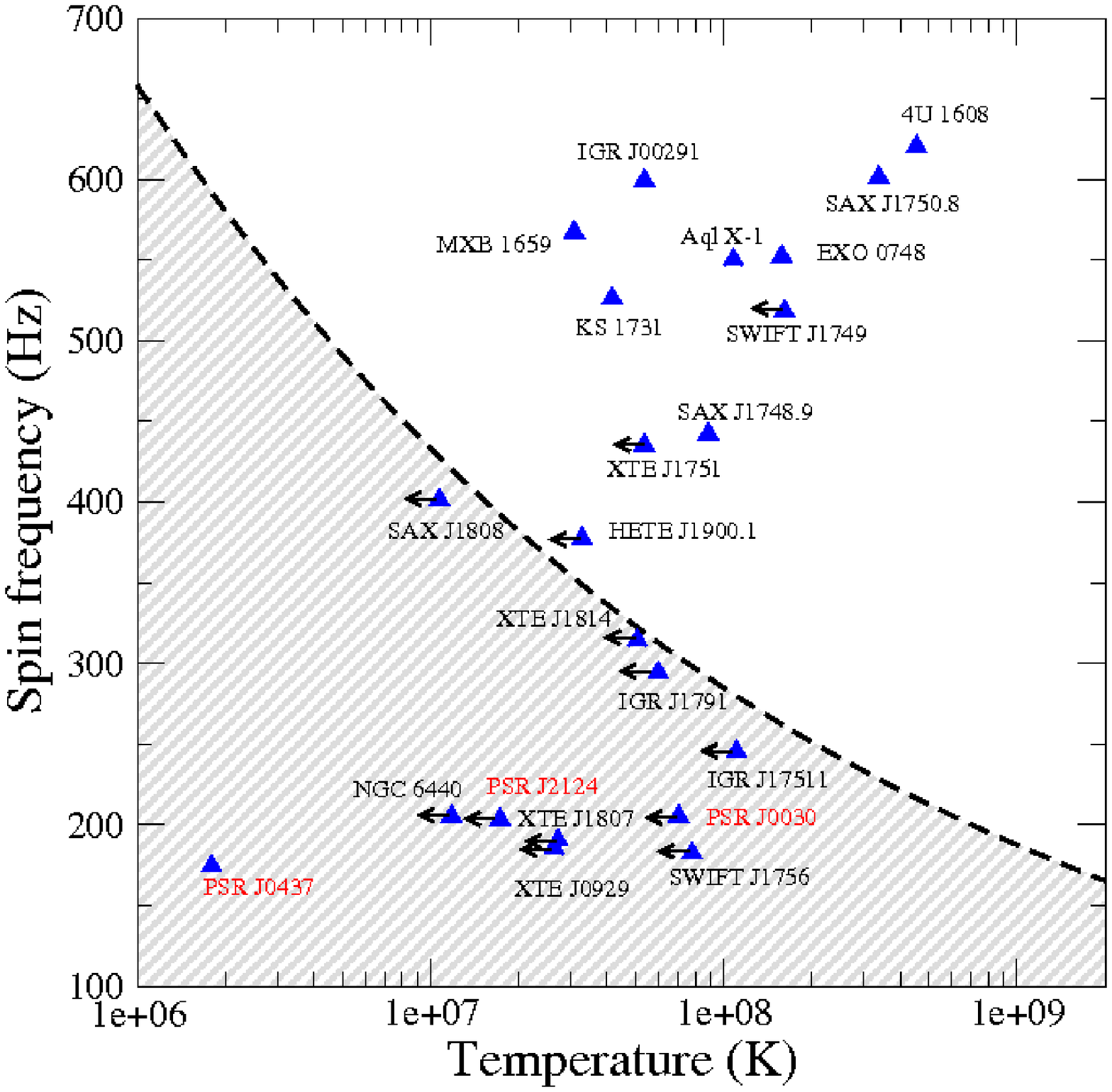}
%
% If no graphics program available, insert a blank space i.e. use
%\picplace{5cm}{2cm} % Give the correct figure height and width in cm
%
\caption{The 'minimal' instability window described in the text, compared to the spin frequency and core temperatures inferred in \cite{HasR} for the known LMXBS. Clearly there are many systems in the instability window, which is inconsistent with the predicted window. To reconcile theory and observations it is thus necessary to either include additional sources of damping in our model. Another possibility is that the r-mode saturation amplitude is small enough that it is indeed unstable, but does not impact on the thermal and spin evolution of this systems.}
\label{window2}
\end{figure}

An analysis by Haskell et al. \cite{HasR}, confirmed by Mahmoodifar and Strohmayer \cite{SiminR}, has however revealed that if one populates the instability window with data from systems where there is both an estimate of the spin period of the NS and its internal temperature (obtained from fits to the surface temperature), many systems would sit inside the instability window, as shown in figure \ref{window2}. This issue has also been examined on a theoretical basis by Ho et al. \cite{HoR}. The conclusion is robust despite the uncertainty introduced by the unknown mass of the star, and the need to model the atmosphere to map the surface temperature to the core (see \cite{HasR} for a detailed description of the different assumptions). It is thus clear that the 'minimal' NS model described above is not consistent with observations. One can also assess the viability of the spin equilibrium scenario by calculating the internal temperature that a star would have if the spin-up torque due to accretion is balanced by an r-mode at the observed spin period. Given the observed accretion luminosity $L_{acc}$, the heat dissipated at equilibrium is \cite{Brown}
\be
L_{heat}=0.064\left(\frac{\nu}{300\mbox{Hz}}\right) L_{acc}.\label{heat}
\ee
The core temperature can be obtained by assuming that the heat in \ref{heat} is carried away by neutrinos. In figure \ref{window2} we show the inferred core temperatures obtained by balancing heating with neutrino emission processes, calculated accounting for modified Urca processes and Copper pair formation, as described in \cite{HoR}. It can be seen that in most cases the stars are too cold to allow for an r-mode to balance the spin-up torque, except for the hotter, faster systems. This is important for GW target selection, as the energy emitted in GWs increases steeply with frequency, making these systems the best targets for next-generation detectors.

Our understanding of the r-mode instability can be made consistent with observations in two ways: either we include additional physics in our models, allowing for additional sources of damping, or we assume that the r-mode saturation amplitude is so small that it has no impact on the thermal and frequency evolution of the star, and that a system can thus 'live' in the instability window.

\begin{figure}[b]
\sidecaption
% For example, with the graphicx style use
\includegraphics[scale=.33]{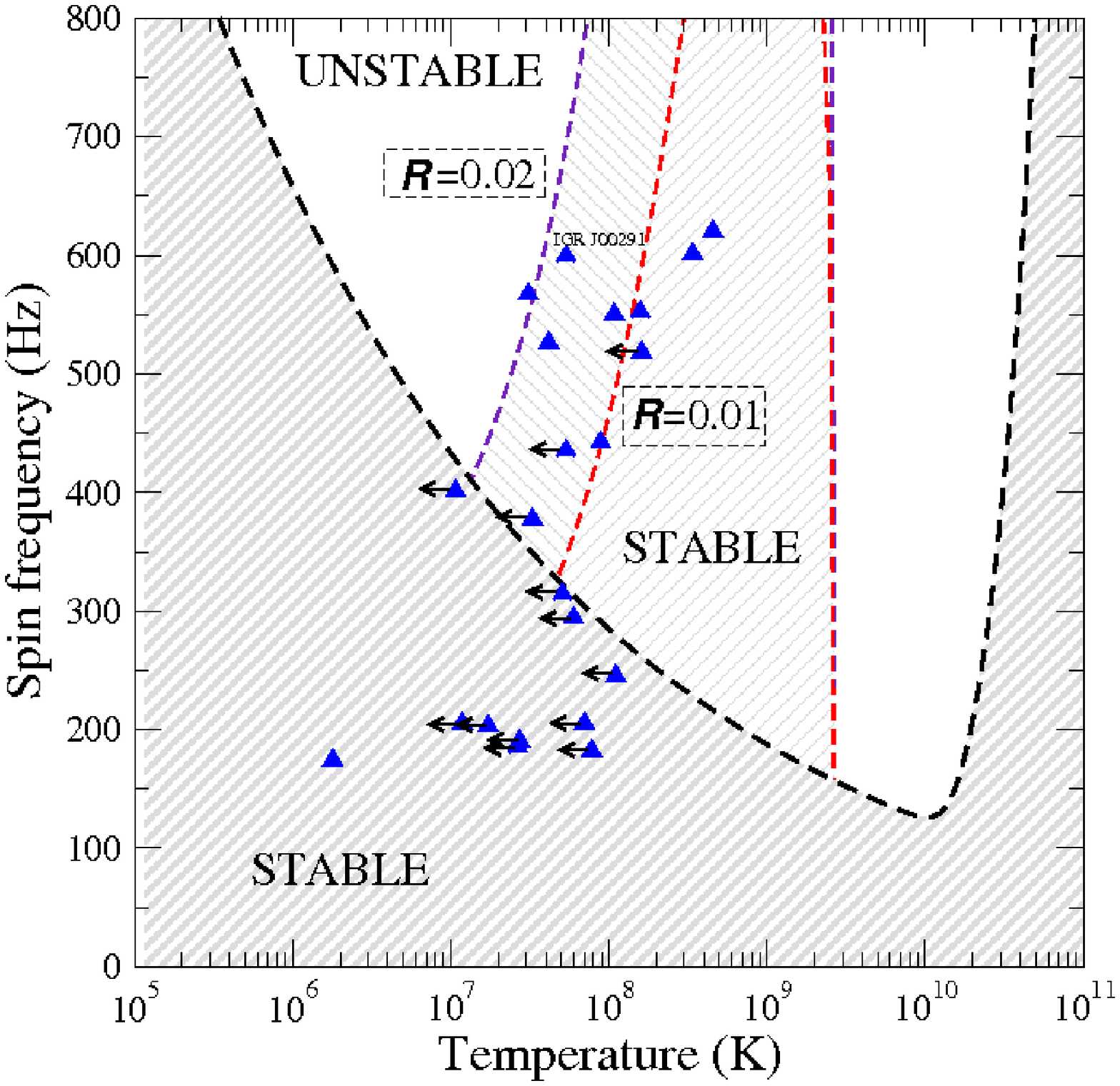}\includegraphics[scale=.33]{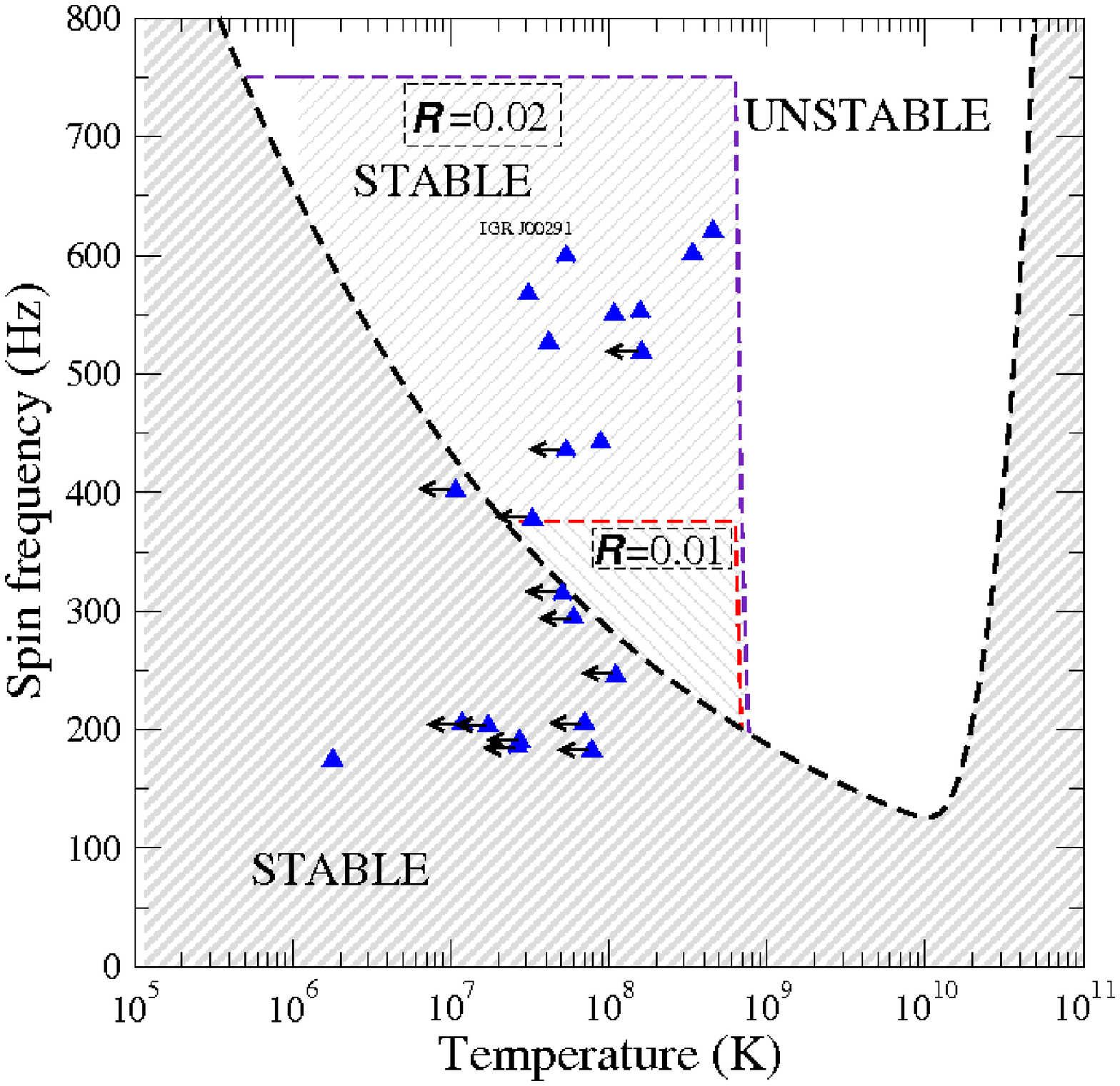}
%
% If no graphics program available, insert a blank space i.e. use
%\picplace{5cm}{2cm} % Give the correct figure height and width in cm
%
\caption{The instability window for strong mutual friction due to vortex/flux tube cutting in the core. We show the results for two different models for the superfluid pairing gaps, the 'strong' (left) and 'weak' (right) models described in \cite{Has09}. In both cases the window can be reconciled with observations for superfluid drag parameters of $\mathcal{R}\approx 0.01$, which is in the possible range for the vortex flux tube cutting mechanism.}
\label{windowMF}
\end{figure}

The first possibility, i.e. that that additional physics modifies the r-mode instability window, has been considered by several authors. Additional viscosity may be provided, among others, by hyperons in the core \cite{HasHyp,BenHyp}, deconfined quarks \cite{HasCore, HasCFL2}, coupling to inertial modes \cite{Gusakov}, torsional oscillations of the crust \cite{HoR} or by magnetic braking \cite{Luciano1, Luciano2}. A particularly interesting possibility is that strong superfluid mutual friction, due to superfluid vortices cutting through magnetic flux tubes in the superconducting interior of the NS, could lead to increased damping. In figure \ref{windowMF} we show the effect on the instability window of assuming a superfluid drag parameter $\mathcal{R}\approx 0.01$, of the order expected if vortices are continuously cutting through flux tubes. The dimensionless parameter $\mathcal{R}$ represents the strength of the mutual friction between the superfluid neutrons and the protons, and couples the two components on a timescale $\tau\approx (1+\mathcal{R}^2)/2\Omega$, with $\Omega$ the rotation frequency of the star. Note that for standard mutual friction, due to electron scattering off magnetised vortex cores, the drag parameter is much smaller, $\mathcal{R}\approx 10^{-4}$, and the effect on the instability window is negligible \cite{Has09,Passa09}.
In the estimates in figure \ref{windowMF}, we have, however, assumed that vortices are free to cut through flux tubes. This may not be the case, as the energy cost associated with cutting effectively 'pins' vortices to flux tubes until a sufficient lag builds up between the neutron an proton fluid, and classical hydrodynamical lift forces, Magnus forces, can push vortices out. This provides a non-linear saturation mechanism for the r-mode, as the mode can only grow to the point where the velocity perturbation is large for the Magnus force to push vortices through flux-tubes (more specifically it is the counter-moving component of the velocity perturbation that grows, but as it grows at the same rate as the total velocity perturbation, this complication can be avoided in the following discussion. See \cite{Has09} for a detailed analysis). At this point the process is strongly dissipative and rapidly damps the mode, thus setting a saturation amplitude $\alpha_s$, which takes the form \cite{HasSat}:
\be
\alpha_s\approx 10^{-6} \left(\frac{\lambda_0}{0.1}\right)^{-1}\left(\frac{\nu}{500\mbox{Hz}}\right)^{-3}\left(\frac{B}{10^8\mbox{G}}\right)^{1/2},
\ee
where $\nu$ is the spin frequency of the star, and $\lambda_0$ is the ratio between the amplitude of the counter moving component of the mode to the amplitude of the oscillation in the total velocity, as described in \cite{HasSat}.

A similar effect could be at work in the deep core, if there is a transition to quark matter. In this case a large enough velocity perturbation could lead to strong bulk viscosity due to the different reactions on the two sides of the interface, which saturates the mode \cite{AlfordSat}. Another possibility is that, if viscosity is weak at the crust-core interface (due e.g. to the presence of so called 'pasta' phases \cite{Ravenhall}), non linear couplings saturate the mode at very low amplitudes \cite{IRA2}. In all these cases the saturation amplitude $\alpha$ could be low enough to allow systems to be r-mode unstable, without any observable signature. In this scenario old systems such as LMXBs are unlikely to lead to strong GW emission due to unstable r-modes, with young NSs being a much more promising GW source \cite{AlfordSchw1,AlfordSchw2}.

\section{Mountains on neutron stars}
\label{sec:mount}

Let us now move on to discuss 'mountains'. As already mentioned the crust of the NS can sustain shearing and a sizeable mountain. The leading order contribution to GW emission will come from the mass quadrupole $Q_{22}$, and theoretical calculations of the yield point of the crust suggest that it could sustain quadrupoles of up to $Q_{22}\approx 10^{39}-10^{40}$ g cm$^2$ \cite{HasMount,Ben}, depending on the mass of the star. This is more than enough to allow for torque balance in LMXBs, which requires \cite{UCB}:
\beq
Q_{eq}&=&3.5\times 10^{37} \left(\frac{M}{1.4 M_\odot}\right)^{1/4}\left(\frac{R}{10^6\mbox{cm}}\right)^{1/4} \left(\frac{\dot{M}}{10^{-9} M_\odot/ \mbox{yr}}\right)^{1/2}\nonumber\\
&&\times \left(\frac{300\mbox{Hz}}{\nu}\right)^{5/2} \mbox{g cm$^2$},
\eeq
where $\dot{M}$ is the average mass accretion rate. It is, however, necessary to understand which physical mechanisms will be at work in a real system, and whether they would allow for a 'maximal' mountain to build up in an LMXB. To address this problem we consider the two main mechanisms that have been suggested: thermal mountains and magnetic mountains.

\subsection{Thermal mountains}

\begin{figure}[b]
\sidecaption
% For example, with the graphicx style use
\includegraphics[scale=.305]{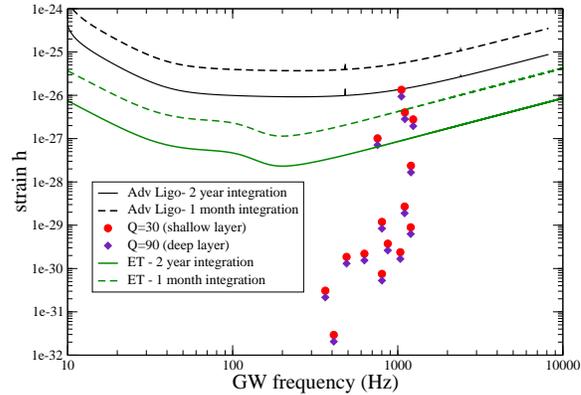}
%
% If no graphics program available, insert a blank space i.e. use
%\picplace{5cm}{2cm} % Give the correct figure height and width in cm
%
\caption{The LIGO and ET sensitivity curves compared to the GW amplitude estimated by Haskell et al. \cite{HasPrep} for transient LMXBs. We can see that assuming that the main contribution is from reactions in a deep or shallow layer makes little difference (i.e. with larger or smaller thresholds $Q$ for the reactions). We show results for both 1 month and 2 year integrations with both detectors, although the maximum time one can track the signal for will be set by the outburst duration, and in most cases will be closer to a month.}
\label{NP}
\end{figure}

\begin{figure}[b]
\sidecaption
% For example, with the graphicx style use
\includegraphics[scale=.302]{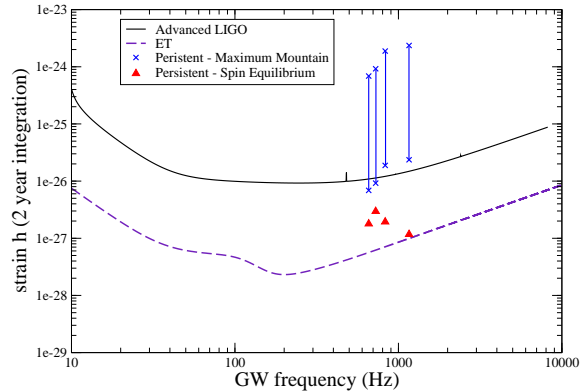}
%
% If no graphics program available, insert a blank space i.e. use
%\picplace{5cm}{2cm} % Give the correct figure height and width in cm
%
\caption{The LIGO and ET sensitivity curves compared to the GW amplitude estimated by {\bf Haskell} for persistent LMXBs. We assume the maximum deformation that the crust can sustain and the error bars account for uncertainties in mass and equation of state. For comparison we also show the GW amplitude that would be needed for torque balance. The deformation needed for spin equilibrium, in fat, smaller than the maximal mountain, so if persistent LMXBs are emitting GWs at the maximal level they will be spinning down.}
\label{Per}
\end{figure}

In the the thermal case the mountain arises as the crust is heated by reactions that occur as accreted material is submerged deep into the crust. As it reaches higher densities several pycno-nuclear reactions occur, which heat the star locally by an amount \cite{Rut}:
\be
\delta T\approx 10^3 C_k^{-1}\left(\frac{p_d}{10^{30}\mbox{erg cm$^{-3}$}}\right)^{-1} \left(\frac{Q_n}{\mbox{MeV}}\right)\left(\frac{\Delta M}{10^{22}\mbox{g}}\right) \mbox{ K},\label{q1}
\ee
where $C_k$ is the heat capacity per baryon in units of the Boltzman constant, $p_d$ is the pressure at which the reaction occur and $Q_n$ the deposited heat per unit baryon. $\Delta M$ is the amount of mass that is accreted. If part of this heating is asymmetric, and quadrupolar in particular, this can lead to a mass quadrupole \cite{UCB}
\be
Q_{22}\approx 1.3\times 10^{35} \left(\frac{R}{10^6\mbox{cm}}\right)^4 \left(\frac{Q}{30 \mbox{MeV}}\right)^3 \left(\frac{\delta T_q}{10^5\mbox{K}}\right) \mbox{ g cm$^2$},\label{q2}
\ee
where $\delta T_q$ is the quadrupolar component of the temperature variation due to the reactions and $Q$ the threshold energy for different reactions. Combining equations (\ref{q1}) and (\ref{q2}), we can thus estimate how large a quadrupole can be built up during an outburst of a certain system, given the observed accretion rate and outburst duration. In figure \ref{NP} we show the results of \cite{HasPrep} for known LMXB transients, that swing from accretion outburst to periods of quiescence. We plot the expected gravitational wave amplitude:
\be
h=\frac{256}{5}\left(\frac{\pi^5}{3}\right)^{1/2}\frac{GQ_{22}\nu^2}{d c^4},
\ee
with $G$ the gravitational constant, $c$ the speed of light and $d$ the distance to the source.
Given that the thermal timescale of the crust is generally quite short (of order a few years for the very deep crust) compared to the quiescence timescale, we assume that in this case the deformation is washed away in-between outbursts, and has to be rebuilt each time. Comparing our results to LIGO and ET sensitivities show that detecting this kind of emission will be very challenging, as also discussed in \cite{Watts08}. The situation is more promising for persistent systems, which undergo long periods of accretion, as illustrated in figure \ref{Per}. Here we have assumed that it is possible to build a 'maximal' mountain in the crust. Even accounting for uncertainties in mass and equation of state, it is clear that these are the most promising targets for Advanced LIGO and ET. Note that it may be possible for compositional asymmetries to persist in accreting systems even in quiescence \cite{UCB}, leading to larger quadrupoles that those estimated by Haskell et al. \cite{HasPrep}. Finally it is also possible for the core to sustain shearing, if it contains a condensate of quarks in the CFL phase \cite{HasCore, BenCore}, although in this case the deformations would be much larger and in some cases can start to be constrained by current LIGO upper limits \cite{LIGO}.

\subsection{Magnetic mountains}

\begin{figure}[b]
\sidecaption
% For example, with the graphicx style use
\includegraphics[scale=.305]{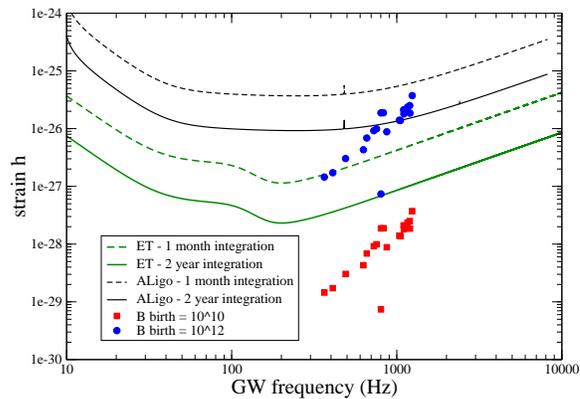}
%
% If no graphics program available, insert a blank space i.e. use
%\picplace{5cm}{2cm} % Give the correct figure height and width in cm
%
\caption{Prediction by Haskell et al. \cite{HasPrep} for the GW emission from known LMXBs, given a magnetic mountain with a background magnetic field of $B=10^{10}$ G or $B=10^{12}$ G. We consider the case in which the mountain is stable in-between outbursts and can thus be built gradually over the life time of the system. Comparing to the sensitivity of Advanced LIGO and ET we can see that only the somewhat extreme case of a background (buried) magnetic field of $B=10^{12}$ G would lead to a detectable signal. }
\label{mount1}
\end{figure}

\begin{figure}[b]
\sidecaption
% For example, with the graphicx style use
\includegraphics[scale=.302]{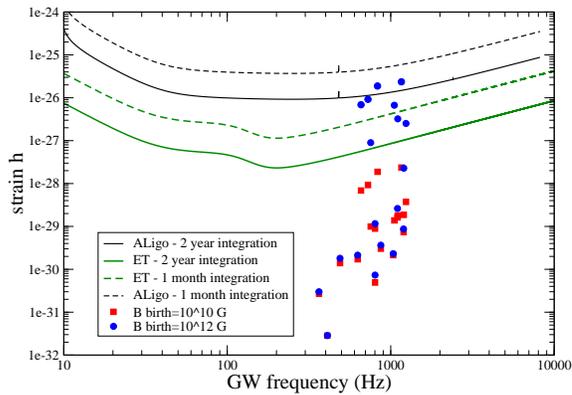}
%
% If no graphics program available, insert a blank space i.e. use
%\picplace{5cm}{2cm} % Give the correct figure height and width in cm
%
\caption{Prediction by Haskell et al. \cite{HasPrep} for the GW emission from known LMXBs, given a magnetic mountain with a background magnetic field of $B=10^{10}$ G or $B=10^{12}$ G. We consider the case in which the mountain is unstable in-between outbursts and is thus dissipated in quiescence. Comparing to the sensitivity of Advanced LIGO and ET we can see, as in the previous case, that only the somewhat extreme case of a background (buried) magnetic field of $B=10^{12}$ G would lead to a detectable signal. }
\label{mount2}
\end{figure}

Accretion does not only lead to thermal perturbations in the crust, but also perturbs the magnetic field structure. After matter is accreted at the magnetic poles it spreads towards the equator, compressing the field and leading to an overall suppression of the large-scale dipolar structure, but also to local enhancements that can support a sizeable mountain \cite{Andrew1,Andrew2,VIG2}. Given an amount of accreted mass $M_a$, the mass quadrupole is given by \cite{Shib, Max1}:
\be
Q_{22}=10^{45}A \left(\frac{M_a}{M_\odot}\right)\left(1+\frac{M_a}{M_c}\right)^{-1}
\ee
where $A\approx1$, $M_c$ is the critical mass at which the process saturates, and the exact value of both quantities is equation of state dependent. In figures \ref{mount1} and \ref{mount2} we show the results obtained by Haskell et al. \cite{HasPrep} for the model E equation of state of Priymak et al. \cite{Max1}. The critical mass depends on the assumed background field of the systems and we consider two possibilities, a background field of $B=10^{12}$ G and one of $B=10^{10}$ G. The latter may be a more realistic limit, given that Grad-Shafranov simulations show that the external dipole can be quenched by approximately an order of magnitude \cite{Max1}, and the inferred dipolar fields of LMXBs and MSRPs are generally in the $B\approx10^8-10^9$ G range.

Dynamical MHD simulations generally confirm stability of the mountain on timescales of $\tau\geq 10^8$ yrs \cite{VIG09} and thus allow to construct the mountain over several outbursts, as assumed in figure \ref{mount1}. Nevertheless, given that MHD simulations may fail to resolve certain instabilities due to finite grid size, we also present the case in which the mountain is dissipated between outbursts in figure \ref{mount2}. In both cases the detection prospects are quite pessimistic, with a detection likely only for the somewhat more extreme hypothesis of a buried $10^{12}$ G magnetic field. 

If a GW signal were to be detected, it would thus have to be from a strong field system and one would expect to see cyclotron features in the electromagnetic emission of the NS, as compared to simple thermal asymmetries in the thermal mountain case. This would offer the possibility, given a GW detection of a continuous signal, to distinguish between the two kinds of NS mountains \cite{HasPrep, MaxPrep}. Note, however, that no cyclotron lines have been detected to date from the known LMXBs containing rapidly rotating NSs, and that the magnetic fields that are inferred for these systems are generally much weaker, of the order of $B\approx 10^8$ G.

\section{Torque balance revisited}

Let us reconsider in more detail the idea that a GW braking torque is needed in LMXBs to balance the accretion spin-up torque. White and Zhang \cite{WZ} argue that the observed cutoff in the frequency distribution of LMXBs and MSRPs could only be explained by the disc/magnetosphere interaction by invoking an unexpected correlation between the magnetic field strength $B$ and mass-accretion rate $\dot{M}$, and assuming that in the more luminous LMXBs the magnetic field strength is greater than what is typically observed in MSRPS (i.e. fields in the range $B\approx10^9-10^{10}$ G would be required). This led to GWs being suggested as a mechanism to remove angular momentum and brake the NS.

However the estimates presented above show that in many systems it would be quite challenging to build a large enough quadrupole with known mechanisms. Furthermore Haskell and Patruno \cite{HasPat} recently considered in detail two systems: SAX J1808.4-3658 and XTE J1814-338. Both these systems are interesting, as the timing solution during an outburst suggests that the frequency is constant \cite{Pat18a,Pat18b}, which is at odds with the theoretical expectation that they should be spinning up due to accretion. This opens up the intriguing possibility that this may be a direct consequence of GWs removing angular momentum from these two systems. Haskell and Patruno \cite{HasPat} considered various GW mechanisms in detail but found that none of them could lead to a large enough quadrupole to balance the spin up torque. In all cases the outbursts are not very luminous and quite short, and SAX J1808.4-3658 has an inferred external dipolar magnetic field of $B\approx 10^8$ G. It is thus unlikely that a mountain, either thermal of magnetic, could provide the necessary quadrupole. Furthermore both systems are too cold to allow for the presence of a large mode of oscillation, such as an r-mode.

However, if one considers a slightly more sophisticated disc model than that used by \cite{WZ} , such as those in \cite{Car1,And05,Car2}, which also account for magnetic torques and radiation pressure, it can be seen that the average accretion rate during the outbursts is actually quite close to the value required for torque balance. In this case, i.e. close to torque balance, the accretion torque is much weaker and can thus account for the absence of a significant increase in spin during the outbursts of these systems. This conclusion is further strengthened by the observation of 1 Hz Quasi-Periodic Oscillations  (QPOs) during reflaring activity at the end of the outburst, which are likely to signal the onset of a propeller phase \cite{PatWatts, ABCPHD, PD}, which has also been observed in other systems \cite{IGR}.

\begin{figure}[b]
\sidecaption
% For example, with the graphicx style use
\includegraphics[scale=.32]{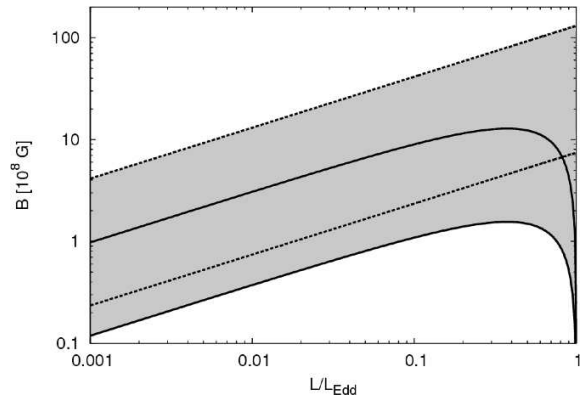}
%
% If no graphics program available, insert a blank space i.e. use
%\picplace{5cm}{2cm} % Give the correct figure height and width in cm
%
\caption{The shaded region represents the region of parameter space allowed by current disc models, in the magnetic field vs luminosity (scaled to the Eddington luminosity) plane, for a hypothetical 730 Hz accreting neutron star at spin equilibrium, as described in \cite{ABCPHD}. As we can see the uncertainties are large enough that the disc/magnetosphere interaction can lead to spin equilibrium at all luminosities for a magnetic field in the range $B\approx 10^8$ G.}
\label{range}
\end{figure}

White and Zhang \cite{WZ} also included in their analysis a number of systems for which the spin frequency had been inferred from the separation between kilohertz QPOs. However it has since been shown that this might not be good proxy for the spin frequency of the star \cite{MB07,Yin07}. A recent analysis of the disc/magnetosphere spin balance scenario has found no correlation between $B$ and $\dot{M}$ for the current sample of LMXBs \cite{ABCPHD}.

In figure \ref{range} we also show the parameter space that is consistent with current accretion disc models, for a system at equilibrium at $\nu=730$ Hz. It is clear that for all accretion rates one can account for the current spin period with a magnetic field in the range of $B\approx 10^8$ G, which is consistent with the values that are inferred from the spin down of MSRPs and of accreting pulsars that have been timed during multiple outbursts.
In conclusion our analysis shows that GW emission is not needed to explain the observed spin distribution of LMXBs and MSRPs, but these systems may be emitting GWs at lower level that can still be detected by next-generation detectors, such as ET.

\section{Summary}

In this chapter we review several GW emission mechanism that may be at work in LMXBs. First of all we consider the r-mode instability in rapidly rotating NSs. Following the analysis of Haskell et al. \cite{HasR} we show that a 'minimal' NS model, that does not include exotica in the core or dynamically significant magnetic fields and superfluid degrees of freedom, is not consistent with the inferred spins and temperatures of NSs in LMXBs. It is thus necessary to include additional physics in our model to account either for additional viscosity that stabilises the mode, or for a very small saturation amplitude that allows the system to be r-mode unstable without any observational impact on its spin and thermal evolutions.  Furthermore most systems are too cold for the torque balance scenario, except for the faster, hotter systems, that may be good targets for next-generation GW detectors such as Advanced LIGO or ET.

We also consider the possibility of a thermal or magnetic mountain being built up during an accretion outburst. We find that persistent systems are a promising GW source, as they would allow to build up a large mountain that could be detected by Advanced LIGO or ET, while for transient systems the mountain is dissipated in quiescence, leading to much lower level emission \cite{HasPrep}. In the magnetic case we find that the GW signal would only be detectable if the buried magnetic field is of the order of $B\approx 10^{12}$ G. 

Finally we re-assess the idea that GWs are needed to provide a braking torque that can balance the spin-up torque due to accretion, and explain the observed spin distribution of LMXBs and MSRPs. We show that current data is consistent with the disc/magnetosphere interaction being the physical mechanism responsible for the observed distribution (although additional mechanisms, such as spin glitches, may be at work in individual systems, see e.g. \cite{WynnG}). Furthermore current disc models allow for systems to be at equilibrium with a magnetic field of $B\approx 10^8$ G. There is thus no need to invoke GWs as a necessary mechanism to provide spin equilibrium in LMXBs, but we have shown that many systems may still be emitting GWs, at a possibly lower level than that required for torque balance. They may thus be interesting sources for next-generation detectors such as the Einstein Telescope.

\begin{acknowledgement}
BH acknowledges the support of the Australian Research Council via a Discovery Early Career Award (DECRA) fellowship. A.P. acknowledges support from the Netherlands Organization for Scientific
Research (NWO) Vidi fellowship
\end{acknowledgement}
%
%\section*{Appendix}
%\addcontentsline{toc}{section}{Appendix}
%
%
%When placed at the end of a chapter or contribution (as opposed to at the end of the book), the numbering of tables, figures, and equations in the appendix section continues on from that in the main text. Hence please \textit{do not} use the \verb|appendix| command when writing an appendix at the end of your chapter or contribution. If there is only one the appendix is designated ``Appendix'', or ``Appendix 1'', or ``Appendix 2'', etc. if there is more than one.

%\begin{equation}
%a \times b = c
%\end{equation}

%%%%%%%%%%%%%%%%%%%%%%%% referenc.tex %%%%%%%%%%%%%%%%%%%%%%%%%%%%%%
% sample references
% %
% Use this file as a template for your own input.
%
%%%%%%%%%%%%%%%%%%%%%%%% Springer-Verlag %%%%%%%%%%%%%%%%%%%%%%%%%%
%
% BibTeX users please use
% \bibliographystyle{}
% \bibliography{}
%
%\biblstarthook{References may be \textit{cited} in the text either by number (preferred) or by author/year.\footnote{Make sure that all references from the list are cited in the text. Those not cited should be moved to a separate \textit{Further Reading} section or chapter.} The reference list should ideally be \textit{sorted} in alphabetical order -- even if reference numbers are used for the their citation in the text. If there are several works by the same author, the following order should be used: 
%\begin{enumerate}

\end{document}